\documentclass[12pt]{article}
\usepackage{amssymb}
\usepackage{amsfonts}
\usepackage{graphicx}
\usepackage{psfrag}
\usepackage{epsfig}
\usepackage{bbm}
\usepackage{axodraw}

\newenvironment{appendletterA}
 {
  \typeout{ Starting Appendix \thesection }
  \setcounter{equation}{0}
  
 }{
  \typeout{Appendix done}
 }
\newenvironment{appendletterB}
 {
  \typeout{ Starting Appendix \thesection }
  \setcounter{equation}{0}
  
 }{
  \typeout{Appendix done}
 }

\newcommand{\inse}{{\Huge $\times$}}

\newcommand{\ini}{\begin{equation}}
\newcommand{\barray}{\begin{eqnarray}}
\newcommand{\fin}{\end{equation}}
\newcommand{\earray}{\end{eqnarray}}

\newcommand{\bsli}{\begin{slide}}
\newcommand{\esli}{\end{slide}}
\newcommand{\bcen}{\begin{center}}
\newcommand{\ecen}{\end{center}}

\newcommand{\bite}{\begin{itemize}}
\newcommand{\eite}{\end{itemize}}
\newcommand{\bmath}{\begin{displaymath}}

\newcommand{\emath}{\end{displaymath}}

\newcommand{\ok}{\omega_k}
\def\reff#1{(\protect\ref{#1})}
\newcommand{\fpo}{$\Delta_F(0)$}
\newcommand{\df}{\Delta_F}
\newcommand{\tll}{{t^\lambda}_\lambda}
\newcommand{\tmn}{t_{\mu \nu}}
\newcommand{\mn}{{\mu \nu}}
\def\slash#1{\setbox0=\hbox{$#1$}#1\hskip-\wd0\hbox to\wd0{\hss\sl/\/\hss}}
\def\slashb#1{\setbox0=\hbox{$#1$}#1\hskip-\wd0\dimen0=5pt\advance
       \dimen0 by-\ht0\advance\dimen0 by\dp0\lower0.5\dimen0\hbox
         to\wd0{\hss\sl/\/\hss}}
\newcommand{\uplr}{\stackrel{\leftrightarrow}}
\newcommand{\fhi}{\varphi}

\begin{document}

\hyphenation{re-nor-ma-li-za-tion}

\begin{flushright}
hep-ph/0305050
\end{flushright}
\vspace{0.5cm}
\begin{center}
{\Large \bf Considerations Concerning the Contributions of Fundamental
 Particles to the Vacuum Energy Density}\\
\vspace{0.7cm}
\vspace{0.5cm}
{\large 
 Giovanni Ossola\footnote{e-mail: go226@nyu.edu} and 
 Alberto Sirlin\footnote{e-mail: alberto.sirlin@nyu.edu}}

\vspace{1.5cm}
{\it Department of Physics, New York University,\\
4 Washington Place, New York, NY 10003, USA.} \\
\vspace{1.5cm}
\end{center}

\bigskip

\begin{center}
\bf Abstract 
\end{center}
The covariant regularization of the contributions of fundamental particles to the vacuum 
energy density is implemented in the Pauli-Villars, dimensional regularization, 
and Feynman regulator frameworks. Rules of correspondence between dimensional regularization
and cutoff calculations are discussed. Invoking the scale invariance of free field theories
in the massless limit, as well as consistency with the 
rules of correspondence, it is argued that quartic
divergencies are absent in the case of free fields, while it is shown that they
arise when interactions are present.

\newpage

\section{Introduction}

It has been pointed out by several authors that one of the most glaring contradictions 
in physics is the enormous mismatch between the observed value of the cosmological constant
and estimates of the contributions of fundamental particles to the vacuum 
energy density \cite{v1}. Specifically, the observed
vacuum energy density in the universe is approximately $0.73 \rho_c$, where 
$\rho_c=3H^2/8\pi G_N\approx 4 \times 10^{-47} \mbox{GeV}^4$ is the critical density, 
while estimates of the contributions of fundamental particles 
range roughly from $(\mbox{TeV})^4$
in broken supersymmetry scenarios to $(10^{19}\ \mbox{GeV})^4 = 10^{76}\ (\mbox{GeV})^4$
if the cutoff is chosen to coincide with the Planck scale. Thus, there is a mismatch 
of roughly 59 to 123 orders of magnitude!

The aim of this paper is to discuss the nature of these contributions by means of 
elementary arguments.

In the case of free particles, it is easy to see that a 
covariant regularization is needed, which we implement in the Pauli-Villars (PV) \cite{v2,v3},
dimensional regularization (DR) \cite{v4}, and Feynman regulator (FR) \cite{v3} frameworks.

We recall that the vacuum energy density is given by
\ini \label{e1}
\rho = <0|T_{00}|0>\, ,
\fin
where $T_{\mu \nu}$ is the energy-momentum tensor.

Defining $t_{\mu \nu} \equiv <0|T_{\mu \nu}|0>$
and assuming the validity of Lorentz invariance,
we have 
\ini \label{e2}
t_{\mu \nu} = \rho\ g_{\mu \nu}\, ,
\fin
or, equivalently,
\ini \label{e3}
t_{\mu \nu} = \frac{g_{\mu \nu}}{4}\ \tll\, ,
\fin
where we employ the metric $g_{00}=-g_{11}=-g_{22}=-g_{33}=1$.

We first consider the case of a free scalar field. Expanding the fields in plane waves 
with coefficients expressed in terms of creation and annihilation operators, and using 
their commutation relations, one readily finds the familiar
expression
\ini \label{e4}
\rho = t_{00} = \frac{1}{2} \int \frac{d^3 k}{(2\pi)^3}\ \frac{\ok^2}{\ok} \, ,
\fin
where $\ok = [(\vec k)^2 + m^2]^{1/2}$, as well as
\ini \label{e5}
p = t_{ii} = \frac{1}{2} \int \frac{d^3 k}{(2\pi)^3}\ \frac{(k_i)^2}{\ok} \, .
\fin
In Eq.~\reff{e5}, $i=1,2,3$ and there is no summation over $i$.

Both  Eq.~\reff{e4} and  Eq.~\reff{e5} are highly divergent and therefore mathematically
undefined. Moreover, as recently emphasized by E.~Kh.~Akhmedov \cite{v5}, the usual
procedure of introducing a three-dimensional cutoff leads to an obvious contradiction:
since the integrands in Eqs.~(\ref{e4}, \ref{e5}) are positive, one would reach the conclusion
that $t_{00}$ and $t_{ii}$ have the same sign, in contradiction with Eq.~\reff{e2}!
This reflects the fact that a three-dimensional cutoff breaks Lorentz invariance. Clearly,
covariant regularization procedures are required!

This is an important issue, since expressions that are not properly regularized
are often deceptive. A classical example is provided by the calculation of
vacuum polarization in QED, in which a quadratically divergent contribution
turns out to be zero upon the imposition of electromagnetic current conservation \cite{v5b}.
Similarly, the same requirement transforms linearly divergent contributions
to the triangle diagrams into convergent ones \cite{v5c}. In fact,
it is important that the regularization procedure respects the symmetries and
partial symmetries of the underlying theory.

The plan of the paper is the following. In Section 2, we discuss rules of correspondence between
the position of the poles in DR and cutoff calculations. In Section 3 
we implement the covariant regularization of $t_{\mu \nu}$ in the PV, DR, and FR frameworks,
starting from Eqs.~(\ref{e2}, \ref{e4}, \ref{e5}). In Section 4 we consider the evaluation of 
$\tll$ on the basis of well-known expressions for the vacuum expectation value
of products of free-field operators, as well as 
Feynman diagrams. Throughout the paper 
the role played by the scale invariance of free field theories in the massless
limit is emphasized. Section 5 illustrates the important effects arising
from interactions by means of two specific examples. Section 6 presents the conclusions. 
Appendix A proves a general theorem concerning the signs 
of the vacuum energy density contributions
of a free scalar field when it is regularized in the PV framework with the minimum number
of regulator fields, while Appendix B illustrates the rules of correspondence in the evaluation
of the one-loop effective potential. Section 4 contains a phenomenological
update of the Veltman-Nambu sum rule for $m_H$\cite{v6,v7} and of an alternative relation
discussed in Ref.~\cite{v8}. 

\section{Rules of Correspondence}

Since DR does not involve cutoffs explicitly, this approach is seldom employed 
in discussions concerning the cosmological constant and hierarchy problems. However,
as it will be shown, it does give valuable information about the nature of the ultraviolet
singularities. Furthermore, it has other important virtues for the problems under consideration:
it respects the scale invariance of free-field theories in 
the massless limit, does not involve unphysical regulator fields, and it is relatively
easy to use in the two-loop calculation carried out in Section 5.

In order to discuss the position of the poles corresponding to specific ultraviolet
divergencies in multi-loop calculations, it is convenient to multiply each
$n$-dimensional integration $\int d^n k$ by $\mu^{4-n}$, where $\mu$ is the
't~Hooft scale. This ensures that the combination of the prefactor and the integration
has the canonical dimension 4.

Let us first consider quadratic ultraviolet divergencies. In cutoff calculations, 
aside from physical masses and momenta, such contributions are proportional
to $\Lambda^2$, where  $\Lambda$ is the ultraviolet cutoff. In DR they must be proportional 
to suitable poles multiplied by $\mu^2$, since this is the only available mass
independent of the physical masses and momenta. If $L$ is the number of loops,
we have the condition $(\mu^{4-n})^L = \mu^2$, or $n = 4 - 2/L$. This means that 
quadratic divergencies exhibit poles at $n = 2, 3, 10/3, \ldots$ for 
$L = 1, 2, 3, \ldots$ loop integrals. The same conclusion has been stated long ago by
M.~Veltman \cite{v6}.

It should be pointed out that this is a useful criterion for scalar integrals of the form
\ini \label{e6}
I_{l,m} = i\ \int \frac{d^n k}{(2\pi)^n}\ \frac{(k^2)^l}{(k^2-M^2)^m}\, ,
\fin
where $l$, $m$ are integers $\ge 0$ and for brevity we have not included
the $i\epsilon$ instruction. When $l$ is a negative integer,
there may appear poles at $n = 2$ which correspond to infrared, rather than
ultraviolet singularities. A useful example is provided by the 
relation \cite{v8}:
\ini \label{e7}
\int \frac{d^n k}{k^2}=\int \frac{d^n k}{k^2-M^2}-M^2\int \frac{d^n k}{k^2 (k^2-M^2)}\, .
\fin
As is well known, the l.h.s. is zero in DR. The first integral in the r.h.s. is quadratically
divergent in four dimensions and consequently exhibits an ultraviolet pole at $n = 2$,
while the second one involves a pole at  $n = 2$ arising from the Feynman
parameter integration. This last singularity is related to the fact
that the second integral contains a logarithmic infrared divergence at  $n = 2$.
Thus, in Eq.~\reff{e7} we witness a cancellation between ultraviolet and infrared
poles. As pointed out in Ref.\cite{v8}, in discussing ultraviolet singularities
one should include in that case the contribution from the first integral.

The above discussion can be extended to quartic divergencies. Since, by an analogous argument,
these must be proportional to $\mu^4$, we have the relation
$(\mu^{4-n})^L = \mu^4$, or $n = 4 - 4/L$. Thus, quartic divergencies exhibit poles
at  $n = 0, 2, 8/3, \ldots$ for $L = 1, 2, 3, \ldots$ loop scalar integrals.
Of course, as we will see in a specific example, a quartic divergence may also
arise from the product of two one-loop quadratically divergent integrals,
each of which has a pole at  $n=2$.

In Section 3, we show how these rules permit to establish a correspondence between
DR and cutoffs calculations in one-loop amplitudes.

\boldmath
\section{Regularization of $t_{\mu \nu}$}
\unboldmath

In order to implement a covariant regularization of $t_{\mu \nu}$, we first search for
a four dimensional representation of Eqs.~(\ref{e4}, \ref{e5}).

Inserting the well-known identity
\ini \label{e8}
\frac{1}{2 \ok} = \int^{\infty}_{-\infty} d k_0\ \delta(k^2-m^2)\ \theta(k_0)\, ,
\fin
$k^2 \equiv k_0^2-(\vec k)^2$, in Eqs.~(\ref{e4}, \ref{e5}), we see that $\rho$ and
$p$ are the zero-zero and $i$-$i$ components of the formal tensor
\ini \label{e9}
t_{\mu \nu} = \int \frac{d^4 k}{(2 \pi)^3}\ k_\mu k_\nu\ \delta(k^2-m^2)\ \theta(k_0)\, .
\fin
As we will discuss in detail later on, Eq.~\reff{e9} can be regularized
in the PV and DR frameworks. Since $t_{\mu \nu}$ is proportional to $g_{\mu \nu}$,
in analogy with  Eqs.~(\ref{e2}, \ref{e3}) it follows that
\ini \label{e10}
t_{\mu \nu} = \frac{g_{\mu \nu}}{4}\ m^2  \int \frac{d^4 k}{(2 \pi)^3}\ \delta(k^2-m^2)\ 
\theta(k_0)\, ,
\fin
where we have employed $k^2\ \delta(k^2-m^2) = m^2\ \delta(k^2-m^2)$. Using
Eq.~\reff{e8} in reverse and the identity
\ini \label{e11}
\frac{1}{\ok} = \frac{1}{\pi} \int^{\infty}_{-\infty} d k_0\ \frac{i}{k^2-m^2+i\epsilon}\, ,
\fin
which follows from contour integration, Eq.~\reff{e10} can be cast in the form
\ini \label{e12}
\tmn = \frac{g_{\mu \nu}\ m^2}{4}\ \int \frac{d^4 k}{(2 \pi)^4}  
\frac{i}{k^2-m^2+i\epsilon}\, ,
\fin
which implies
\ini \label{e13}
\tll =  m^2\ \int \frac{d^4 k}{(2 \pi)^4} \frac{i}{k^2-m^2+i\epsilon}\, .
\fin
We note that the integral in Eqs.~(\ref{e12}, \ref{e13}) is $i$\fpo, the Feynman
propagator evaluated at $x=0$. 

An alternative derivation of Eqs.~(\ref{e12}, \ref{e13}) can be obtained by
using the starting Eqs.~(\ref{e4}, \ref{e5}) to evaluate the trace $\tll$:
\ini \label{e14}
\tll =  \frac{1}{2} \int \frac{d^3 k}{(2 \pi)^3} \frac{\ok^2-(\vec k)^2}{\ok} =
\frac{m^2}{2} \int \frac{d^3 k}{(2 \pi)^3} \frac{1}{\ok}\, .
\fin
Combining Eqs.~(\ref{e3}, \ref{e11}, \ref{e14}), we immediately recover
Eqs.~(\ref{e12}, \ref{e13})! It is important to note that these expressions
are proportional to $m^2$ and to the quadratically divergent integral $i$\fpo.

Returning to the issue of regularization, in the PV framework Eq.~\reff{e9}
is replaced by the regularized expression
\ini \label{e15}
(\tmn)_{PV} =\sum_{i=0}^N\ C_i\ \int \frac{d^4 k}{(2 \pi)^3}\ k_\mu k_\nu\ 
\delta(k^2-M_i^2)\ \theta(k_0)\, ,
\fin
where $M_0=m$, $C_0=1$, N is the number of regulator fields, $M_j$ ($j=1,2,\ldots,N$) 
denote their masses, and the $C_i$ obey the constraints
\ini \label{e16}
\sum_{i=0}^N\ C_i\ (M_i^2)^p = 0 \quad \, (p=0,1,2)\, .
\fin
From Eq.~\reff{e16} we see that in our case $N=3$ is the minimum number of regulator fields.
It is worthwhile to note that if the limit $M_j \to \infty$ is taken before the integration
is carried out, Eq.~\reff{e15} reduces to the original, unregularized expression of
Eq.~\reff{e9}. In fact, $\lim_{M_j \to \infty}\ \delta(k^2-M_j^2) = 0$.
Following the steps leading from Eq.~\reff{e9} to Eq.~\reff{e12}, Eq.~\reff{e15} becomes
\ini \label{e17}
(\tmn)_{PV} = \frac{g_{\mu \nu}}{4}\ \sum_{i=0}^N\ C_i\ M_i^2 
\int \frac{d^4 k}{(2 \pi)^4}\ \frac{i}{k^2-M_i^2+i\epsilon}\ ,
\fin
which is the PV regularized version of Eq.~\reff{e12}.

The simplest way to evaluate Eq.~\reff{e17} is to differentiate twice
\ini \label{e18}
I(M_i^2) \equiv \int \frac{d^4 k}{(2 \pi)^4}\ \frac{i}{k^2-M_i^2+i\epsilon}\, ,
\fin
with respect to $M_i^2$, so that it becomes convergent. This leads to 
$I''(M_i^2)=1/16\pi^2 M_i^2$. Integrating twice $I''(M_i^2)$ with respect to  $M_i^2$,
we obtain
\ini \label{e19}
I(M_i^2) = \frac{1}{16\pi^2}\left[ M_i^2\ ( \ln{M_i^2}-1 )+K_1 M_i^2 +K_2\right]\,  ,
\fin
where $K_1$ and $K_2$ are arbitrary constants of integration. When inserted in
Eq.~\reff{e17}, the terms involving $K_1$, $K_2$, and $-M_i^2$ cancel on account of
 Eq.~\reff{e16}, and  Eq.~\reff{e17} becomes
\ini \label{e20}
(\tmn)_{PV} = \frac{g_{\mu \nu}}{64\pi^2}\ \sum_{i=0}^N\ C_i\ M_i^4\ 
\ln{\left(\frac{M_i^2}{\nu^2}\right)}\, . 
\fin
Here $\nu$ is a mass scale that can be chosen arbitrarily since its contribution vanishes
on account of Eq.~\reff{e16} with $p=2$.

If we choose $N=3$, the minimum number of regulator fields, the constants $C_i$
can be expressed in terms of the $M_i$ by solving Eq.~\reff{e16} for $p=0,1,2$
(see Appendix A). One finds that Eq.~\reff{e20} contains three classes of
contributions: i) terms quartic in the regulator masses $M_j$ ($j=1,2,3$);
ii) terms of ${\cal O}(m^2)$ which are quadratic in $M_j$; iii) terms of
${\cal O}(m^4)$. As shown in Eq.~\reff{e20}, all these contributions are accompanied
by logarithms. If one rescales the regulator masses by a common factor $\Lambda$,
one finds that, modulo logarithms, the three classes become proportional to
$\Lambda^4$, $\Lambda^2$, and $\Lambda^0$, respectively.
Thus, the first class of terms exhibit the quartic divergence frequently invoked in
discussions of the cosmological constant problem. However, as shown in the Appendix, the results
show a curious and at first hand unexpected feature: for arbitrary values of $M_j$,
the sign of the quartic contribution to $\rho$ is negative!
Instead, the sign of the ${\cal O}(m^2 \Lambda^2)$ term is positive, and that
of ${\cal O}(m^4)$ contribution is negative.

The PV expression greatly simplifies in the limit $M_3 \to M_2 \to M_1 = \Lambda$
and becomes
\ini \label{e21}
(\tmn)_{PV} = \frac{g_{\mu \nu}}{128\pi^2}\ \left[-\Lambda^4+4 m^2 \Lambda^2 
- m^4 \left(3+2 \ln{\frac{\Lambda^2}{m^2}}\right)\right]\, . 
\fin
It is interesting to note that Eq.~\reff{e17} also follows from the PV regularization
of the formal, quartically divergent tensor
\ini \label{e22}
J_{\mu \nu} = i \int \frac{d^4 k}{(2 \pi)^4}\ \frac{k_\mu k_\nu}{k^2-m^2+i\epsilon}\, .
\fin
In fact
\bmath
(J_{\mu \nu})_{PV} = i \sum_{i=0}^N\ C_i\ \int \frac{d^4 k}{(2 \pi)^4}\ 
\frac{k_\mu k_\nu}{k^2-M_i^2+i\epsilon}\, ,
\emath
which, upon the replacement $k_\mu k_\nu \to g_{\mu \nu} k^2 / 4$ and the
decomposition $k^2 = k^2 -M_i^2 + M_i^2$, reduces to Eq.~\reff{e17},
since the contribution of $ k^2 -M_i^2$ vanishes on account of Eq.~\reff{e16}.
Eq.~\reff{e22} may be also regulated by means of a Feynman regulator which, for
this application, we choose 
to be of the form $[(\Lambda^2 - m^2)/(\Lambda^2 -k^2 -i \epsilon)]^3$. Thus,
\ini \label{e23}
(J_{\mu \nu})_{FR} = i \int \frac{d^4 k}{(2 \pi)^4}\ 
\frac{k_\mu k_\nu}{k^2-m^2+i\epsilon} 
\frac{(\Lambda^2 - m^2)^3}{(\Lambda^2 -k^2 -i \epsilon)^3}\, .
\fin
Evaluating Eq.~\reff{e23} we find the very curious result that it exactly
coincides with Eq.~\reff{e21}!
An advantage of Eq.~\reff{e23} is that one can discern immediately its sign
by means of a Wick rotation of the $k_0$ axis. 
Replacing $k_\mu k_\nu \to g_{\mu \nu} k^2 / 4$, performing the rotation and
introducing $k_0 = i K_0$, we obtain the Euclidean representation
\ini \label{e24}
(J_{\mu \nu})_{FR} = - \frac{g_{\mu \nu}}{4}\ \int \frac{d^4 K}{(2 \pi)^4}\ 
\frac{K^2}{K^2+m^2} 
\frac{(\Lambda^2 - m^2)^3}{(\Lambda^2 +K^2)^3}\, ,
\fin
which shows that the cofactor of $g_{\mu \nu}$ is manifestly negative, in conformity
with Eq.~\reff{e21}.

In summary, according to  Eq.~\reff{e20} with the minimum number $N=3$ of regulator
fields, or its limit in Eq.~\reff{e21}, the leading contribution for a bosonic field
would be $\rho = - {\cal O}(\Lambda^4)$, $p = {\cal O}(\Lambda^4)$!
Such a result is theoretically unacceptable since for a scalar field
\bmath
 <0|T_{00}|0> = \frac{1}{2} <0|\partial_0 \fhi \partial_0 \fhi +
 \partial_i \fhi \partial_i \fhi +m^2 \fhi^2 |0>
\emath
should be positive.
We therefore interpret the sign problem as an artifact of the regularization procedure
that arises in the case $N=3$ due to the fact that some of the regulator fields 
have negative metric.

The presence of quartic divergencies, of either sign, has another highly
unsatisfactory consequence, namely it breaks down the scale invariance of
free field theories in the massless limit! We recall that the divergence
of the dilatation current for a scalar field has the form
\ini \label{e25}
\partial_\mu D^\mu = {T^\lambda}_\lambda + \frac{\square \phi^2}{2} 
= {\Theta^\lambda}_\lambda\,  ,
\fin
where $\Theta_\mn$ is the ``improved'' energy-momentum tensor \cite{v9}.
This leads to
\ini \label{e26}
<0|\partial_\mu D^\mu |0> = \tll = m^2 <0|\phi^2|0>\, ,
\fin
where we used $\square\!\!\!<0|\phi^2|0>=0$ and, in deriving the second equality in
Eq.~\reff{e26}, we employed the equation of motion. Thus, for free fields,
$\tll$ should vanish as $m \to 0$, a property that is violated by quartic divergencies of
either sign.

In order to circumvent the dual problems of sign and breakdown of scale invariance
of the free-field theory 
in the massless limit within the PV framework, there are two possibilities: one
is to subtract the offending  ${\cal O}(\Lambda^4)$ term in Eq.~\reff{e21}; the
other is to employ $N \ge 4$, in which case the $C_i$ are not determined by
the $M_i$, and the sign of the ${\cal O}(\Lambda^4)$
contributions is undefined. In the last approach one can in principle impose
the cancellation of the quartic divergence as a symmetry requirement.
However, the sign of the leading ${\cal O}(m^2 \Lambda^2)$ term remains
undefined, which is not an attractive state of affairs.

A simpler and more satisfactory approach is to go back to Eq.~\reff{e12} as a starting point to
implement the regularization procedure. Regularization of Eq.~\reff{e12} in the PV
framework would lead us back to  Eqs.~(\ref{e17}, \ref{e21}).
Instead, we may regularize the integral in Eq.~\reff{e12} with a Feynman regulator,
which we choose to be of the form $[\Lambda^2 /(\Lambda^2 -k^2 -i \epsilon)]^2$.
Neglecting terms of  ${\cal O}(m^2/ \Lambda^2)$,
this leads to
\ini \label{e27}
(\tmn)_{FR} = \frac{g_{\mu \nu}}{64 \pi^2}\ 
\left[ m^2 \Lambda^2 - m^4 \left(\ln{\frac{\Lambda^2}{m^2}} - 1 \right)\right]\, ,
\fin
which implies
\ini \label{e28}
(\tll)_{FR} = \frac{1}{16 \pi^2}\ 
\left[ m^2 \Lambda^2 - m^4 \left(\ln{\frac{\Lambda^2}{m^2}} - 1 \right)\right]\, .
\fin
Eqs.~(\ref{e27}, \ref{e28}) have the correct sign and conform with 
scale invariance in the massless limit! A similar result is obtained if 
a Wick rotation is implemented in Eq.~\reff{e12}, and an invariant cutoff 
is employed to evaluate the integral.

We now turn to DR. Since the steps from Eq.~\reff{e9} to  Eq.~\reff{e12}
involve only the $k_0$ integration, in DR the regularized expressions of
Eq.~\reff{e9} and Eq.~\reff{e12} are equivalent and we obtain
\ini \label{e29}
(\tmn)_{DR} = \frac{g_{\mu \nu}\ m^2\ \mu^{(4-n)}}{n}\ \int \frac{d^n k}{(2 \pi)^n}  
\frac{i}{k^2-m^2+i\epsilon}\, ,
\fin
which leads to
\ini \label{e30}
(\tll)_{DR} =  m^2\ \mu^{(4-n)}\ \int \frac{d^n k}{(2 \pi)^n} 
\frac{i}{k^2-m^2+i\epsilon}\, .
\fin
We note that the $n=0$ pole in Eq.~\reff{e29} arises from the replacement
$k_\mu k_\nu \to g_{\mu \nu} k^2 / n$. Since $g^{\mu \nu} g_{\mu \nu} = n$,
this pole is absent in the evaluation of the trace in Eq.~\reff{e30}.
In order to use the rules of correspondence in an unambiguous manner,
we apply them to the Lorentz scalar $\tll$ evaluated in the FR and DR frameworks.
Carrying out the integration in Eq.~\reff{e30}, we find
\ini \label{e31}
(\tll)_{DR} =  \frac{4\ m^4}{(2\sqrt{\pi})^n} 
\frac{(\mu/m)^{(4-n)}\ \Gamma(3-n/2)}{(2-n)(4-n)}\, .
\fin 
This expression exhibits poles at $n=4$ and $n=2$ which, according to the rules
of correspondence for one-loop integrals, indicate the presence of logarithmic
and quadratic divergencies.

A heuristic way to establish a correspondence between the cutoff calculation
in Eq.~\reff{e28} and the DR expression in Eq.~\reff{e31}, is to carry out
the expansion about $n=4$ in the usual way, but at the same time separate out the $n=2$
pole in such a manner that the overall result is only modified in ${\cal O}(n-4)$.
This leads to {\small
\ini \label{e32}
(\tll)_{DR} =  \frac{\mu^2 m^2}{2 \pi} \left[ \frac{1}{2-n}+\frac{1}{2}\right]
-  \frac{m^4}{16 \pi^2} \left[ \frac{2}{4-n}+ \ln{\frac{\mu^2}{m^2}}-2C+1\right]
+{\cal O}(n-4)\, ,
\fin}
where $C=[\gamma-\ln{4 \pi}]/2$. The contribution proportional to $m^4$ represents
the usual result. The first term contains the pole at $n=2$, and only modifies
the expansion in ${\cal O}(n-4)$. A correspondence with Eq.~\reff{e28} can be implemented by
means of the identifications
\barray \label{e33}
\left[\frac{1}{4-n}+ \ln{\frac{\mu}{m}}-C\right]_{n \approx 4}& \to & 
\ln{\frac{\Lambda}{m}} - 1\, ,\\ \label{e34}
\frac{\mu^2}{2 \pi} \left[\frac{1}{2-n}+\frac{1}{2}\right]_{n \approx 2} 
&\to & \frac{\Lambda^2}{16 \pi^2}\, ,
\earray
where, for instance, $n \approx 2$ means that $n$ is in the immediate neighborhood of 2.
It is interesting to note that if one approaches
the ultraviolet poles from below, as it seems natural in DR,  
the signs of the left and right sides of Eqs.~(\ref{e33}, \ref{e34}) coincide!

Another interesting information contained in the DR expression of 
Eq.~\reff{e31} is that the ${\cal O} (m^2 \Lambda^2)$
contribution is not accompanied by a $\ln{(\Lambda^2/ m^2)}$ cofactor. This can be 
seen as follows: since Eq.~\reff{e31} is proportional to $(m^2)^{n/2}$,
if we differentiate twice with respect to $m^2$ we see that the pole at $n=2$ disappears. 
As a consequence, terms of ${\cal O} (m^2 \Lambda^2 \ln{(\Lambda^2/ m^2)})$
cannot be present,
since otherwise contributions of ${\cal O} (\Lambda^2)$ would survive under the double 
differentiation. Indeed, this observation agrees with Eq.~\reff{e28}.
Thus, in one-loop calculations depending on $m^2,\Lambda^2$, terms of 
${\cal O} (m^2 \Lambda^2 \ln{(\Lambda^2/ m^2)})$ would require a double pole at $n=2$ 
in the DR expression.

A conclusion essentially identical to Eq.~\reff{e27}, namely that the divergence
of the zero-point energy for free particles is quadratic rather then quartic,
and that massless particles don't contribute, has been recently advocated by
E.~Kh.~Akhmedov \cite{v5}, invoking arguments of relativistic invariance.
The analysis of the present paper shows that this is not enough to single out
Eq.~\reff{e27}, since the PV regularization leads to the covariant
expressions of Eqs.~(\ref{e20}, \ref{e21}) that exhibit a quartic divergence.
What singles out Eq.~\reff{e27} are the combined requirements of
relativistic covariance and scale invariance of free-field theories 
in the massless limit, as well as consistency with the rules of correspondence.

We conclude this Section by recalling that the contributions of all bosons
(fermions) carry the same (opposite) sign as Eq.~\reff{e27}. Each
contribution must be multiplied by a factor $\eta$ that takes into account
the color and helicity degrees of freedom, as well as the particle-antiparticle
content.

\boldmath
\section{Evaluation of $\tll$ based on Feynman Diagrams}
\unboldmath

In Section 3 we have discussed the regularization of $\tmn$ and its trace in
the free-field theory case, starting from the familiar expressions
for $\rho$ and $p$ given in  Eqs.~(\ref{e4}, \ref{e5}).
It is instructive to revisit the evaluation of $\tll$ on the basis of 
well known expressions for the vacuum expectation value of products
of free-field operators on the one hand, and Feynman diagrams on the other.
This will also pave the way to the discussion of the effect of interactions
in Section 5.

We will consider three examples: an hermitian scalar field, a spinor field,
and a vector boson, all endowed with mass $m$.
We recall the free-field theory expressions for ${T^\lambda}_\lambda$
in the three cases:
\ini \label{e35b}
{T^\lambda}_\lambda = -\partial_\lambda \fhi \partial^\lambda \fhi + 2 m^2 \fhi^2
\fin
\ini \label{e36b}
{T^\lambda}_\lambda = - 3 \bar{\psi} \left[ i \frac{\uplr{\slash{\partial}}}{2} 
-  m \right] \psi + m \bar{\psi} \psi
\fin
\ini \label{e37b}
{T^\lambda}_\lambda = - m^2 A_\lambda A^\lambda
\fin
The above formulae are valid in four dimensions and, in deriving  Eq.~\reff{e37b},
we have employed the symmetric version of $T_\mn$ for the spin 1 field.

A direct way of evaluating $\tll$ is to consider the vacuum expectation value 
of two fields at $x$ and $y$, carry out the differentiations exhibited in  
 Eqs.~(\ref{e35b}, \ref{e36b}) and take the limit $x \to y$. For instance,
in the free-field scalar case we have the well-known representation:
\ini \label{e38b}
 <0|\fhi(x) \fhi(y)|0> =  \int \frac{d^4 k}{(2 \pi)^3}\ \delta(k^2-m^2)\ \theta(k_0)\
e^{- i k (x-y)}\, .
\fin
The r.h.s. of Eq.~\reff{e38b}, the $i\Delta^{+}(x-y)$ function, 
is the contribution of the one-particle intermediate state in the
K\"allen-Lehmann representation which, of course, is the only one that survives 
in the free-field theory case. 
From  Eq.~\reff{e38b} we find 
\bmath
 <0|-\partial_\lambda \fhi(x) \partial^\lambda \fhi(y) + 2 m^2 \fhi(x) \fhi(y)|0>\  =
\emath
\ini \label{e38bis}
= \int \frac{d^4 k}{(2 \pi)^3}\ \delta(k^2-m^2)\ \theta(k_0)\ (2 m^2 - k^2)\
e^{- i k (x-y)}\, ,
\fin
which is well defined. Taking the limit $x \to y$ and recalling Eq.~\reff{e35b}, we
obtain the formal expression 
\ini \label{e39b}
 (\tll)_\fhi =  \int \frac{d^4 k}{(2 \pi)^3}\ \delta(k^2-m^2)\ \theta(k_0)\
 (2 m^2 - k^2)\, .
\fin
If instead of $T_\mn$, the ``improved'' tensor $\Theta_\mn$ \cite{v9} is employed 
for scalar fields, Eq.~\reff{e35b} is replaced by 
${\Theta^\lambda}_\lambda = \fhi \Box \fhi + 2 m^2 \fhi^2$, which again leads to
Eqs.~(\ref{e38bis}, \ref{e39b}).
If we replace $k^2 \to m^2$ in these expressions on account of $\delta(k^2-m^2)$,
we recover Eq.~\reff{e10}, the result of our previous analysis in Section 3.
Parenthetically, we recall that, at the classical level, use of the equations
of motion leads to ${\Theta^\lambda}_\lambda =  m^2 \fhi^2$ even in the presence
of the $\lambda \fhi^4$ interaction \cite{v9}.

Eq.~\reff{e39b} admits another representation that can be linked with a Feynman vacuum
diagram, to wit
\ini \label{e40b}
 (\tll)_\fhi = \mbox{Re} \int \frac{d^4 k}{(2 \pi)^4}\ 
\frac{i\ [2 m^2 - k^2]}{k^2 - m^2 + i \epsilon}\, ,
\fin
where we have employed $\pi \delta(k^2-m^2) = \mbox{Re}(i/k^2-m^2+i\epsilon)$
and used the fact that the integrand is even in $k_0$ to replace $\theta(k_0) \to 1/2$.

Equivalently, we have 
\ini \label{e41b}
 (\tll)_\fhi = \mbox{Re} \left\{ m^2  \int \frac{d^4 k}{(2 \pi)^4}\ 
\frac{i}{k^2 - m^2 + i \epsilon} - i \int \frac{d^4 k}{(2 \pi)^4} \right\}\, .
\fin
Eq.~\reff{e40b} is depicted in Fig.~1, where the cross indicates the insertion of the operator
 ${T^\lambda}_\lambda$ given in Eq.~\reff{e35b}. We note that the ``Re'' instruction
is important in the passage from Eq.~\reff{e39b} to Eq.~\reff{e40b} and ensures that 
the answer is real, as required for diagonal matrix elements of the hermitian operator
 ${T^\lambda}_\lambda$.

Using Eq.~\reff{e36b}, the corresponding expression in the fermion case is
\ini \label{e42b}
 (\tll)_\psi = - \mbox{Re}\ \mbox{Tr} \int \frac{d^4 k}{(2 \pi)^4}\ 
\frac{i\ [m - 3 (\slash{k}-m)]}{\slash{k} - m + i \epsilon}\, .
\fin
This can be cast in the form
\ini \label{e43b}
 (\tll)_\psi = - \mbox{Re} \left\{4\ m^2  \int \frac{d^4 k}{(2 \pi)^4}\ 
\frac{i}{k^2 - m^2 + i \epsilon} - 12\ i \int \frac{d^4 k}{(2 \pi)^4} \right\}\, ,
\fin
where we have employed $\int d^4 k\ \slash{k}/(k^2 - m^2 + i \epsilon) = 0$.

Finally, using Eq.~\reff{e37b}, we have
\ini \label{e44b}
 (\tll)_A =  \mbox{Re} \left\{3\ m^2  \int \frac{d^4 k}{(2 \pi)^4}\ 
\frac{i}{k^2 - m^2 + i \epsilon} - \ i \int \frac{d^4 k}{(2 \pi)^4} \right\}\, .
\fin

These expressions exhibit interesting features: the first terms in 
 Eqs.~(\ref{e41b}, \ref{e43b}, \ref{e44b}) are quadratically divergent and real.
The reality condition is easily checked by performing the $k_0$ contour integration
or by means of a Wick rotation of the $k_0$ axis accompanied by 
the change of variable $k_0 = i K_0$. This rotation is mathematically allowed
since the $k_0$ integrations in those contributions are convergent.
The second terms in   Eqs.~(\ref{e41b}, \ref{e43b}, \ref{e44b}) formally exhibit
a quartic divergence. However, since the integrations are over the real axes, such 
terms are purely imaginary in Minkowski space and 
therefore do not contribute if the ``Re'' restriction
is imposed. We note parenthetically that, unlike in the previous case, the Wick rotation
cannot be applied to the unregularized $\int d^4 k$ as it stands since, in performing the $k_0$
integration, the contributions of the large quarter circles in the complex plane are not 
negligible and, in fact, they are necessary to satisfy Cauchy's theorem.

In the PV and DR approaches the regularized versions of the imaginary
contributions in  Eqs.~(\ref{e41b}, \ref{e43b}, \ref{e44b}) vanish automatically.
In contrast, a Feynman regulator of the form $[\Lambda^2/(\Lambda^2-k^2 - i \epsilon)]^3$
leads to
\ini \label{e45b}
 - i  \int \frac{d^4 k}{(2 \pi)^4}\ 
\left( \frac{\Lambda^2}{\Lambda^2 - k^2 - i \epsilon} \right)^3 = \frac{\Lambda^4}{32 \pi^2}\, ,
\fin
which is real and positive, and exhibits the frequently invoked quartic divergence. 
The reality property can also be checked by performing a Wick rotation
in Eq.~\reff{e45b}, which is now mathematically allowed. Thus, we see that the quartically
divergent contributions in the one-loop vacuum diagrams have a very ambivalent and
disturbing property: their contributions to $\rho$ are imaginary in Minkowski space
and real, if regularized according to Eq.~\reff{e45b}, in Euclidean space. 

However,
if  Eq.~\reff{e45b} is applied to regularize the imaginary parts between curly brackets in 
Eqs.~(\ref{e41b}, \ref{e43b}, \ref{e44b}), serious inconsistencies emerge.
In fact, their coefficients do not conform with the relations 
$(\tll)_\psi = - 4\ (\tll)_\fhi$ and $(\tll)_A = 3\ (\tll)_\fhi$,
which arise on account of the helicity and particle-antiparticle degrees of freedom 
of Dirac spinors and massive vector bosons in four dimensions.

A direct way to see that these terms are inconsistent with Eq.~\reff{e39b} is 
to go back to that expression, replace $k^2 \to m^2$ on account to the $\delta$-function
and then use $\delta(k^2-m^2) = (1/\pi) \mbox{Re}(i/k^2-m^2+i\epsilon)$.
This leads to the first term of Eq.~\reff{e41b}, a result that is only
consistent with Eq.~\reff{e39b} if the second contribution vanishes.

We conclude that, in order to avoid inconsistencies, the quartically divergent
imaginary parts in Eqs.~(\ref{e41b}, \ref{e43b}, \ref{e44b}) must be subtracted
either by imposing the reality condition in Minkowski space or by means of the regularization
procedure, as in the DR and PV cases. 
The surviving terms in  
Eqs.~(\ref{e41b}, \ref{e43b}, \ref{e44b}) are proportional to $m^2 i$\fpo,
satisfy the relations $(\tll)_\psi = - 4\ (\tll)_\fhi$ and $(\tll)_A = 3\ (\tll)_\fhi$,
and coincide with the result in Eq.~\reff{e39b} and its equivalent expression in 
Eq.~\reff{e10}. Furthermore, they conform with the scale invariance of free field theories in
the massless limit.

The PV, DR, and FR regularizations and their correspondence was discussed in detail
in Section 3 in the case of the free scalar field, starting with 
Eqs.~(\ref{e12}, \ref{e13}). In writing down the rules of correspondence between DR
and four dimensional cutoff calculations in the case of spin 1 and spin $1/2$ fields,
there is a subtlety that should be pointed out. In the case of the spin 1 field, the DR
version of $\tll$ is given by the expression for the scalar field (Eq.~\reff{e30})
multiplied by $n-1$, the number of helicity degrees of freedom in $n$ dimensions. In separating
out the contribution of the $n=2$ pole (Cf.~Eq.~\reff{e32}), the residue
carries then a factor 1, rather than 3. In order to maintain the proper relation
with the four dimensional calculation,  in the spin 1 case the l.h.s. of Eq.~\reff{e34}
corresponds to $3 \Lambda^2 / 16 \pi^2$ rather than $\Lambda^2 / 16 \pi^2$,
the factor 3 reflecting the number of helicity degrees of freedom in four dimensions.
A similar rule holds for spin $1/2$ fields: if in evaluating the $n=2$ residue
one employs $\mbox{Tr}\ \mathbbm{1} = 2$, as befits a spinor in two-dimensions,
in the rule of correspondence with the four-dimensional cutoff
calculation one includes an additional factor 2 to reflect the fact that 
 $\mbox{Tr}\ \mathbbm{1} = 4$ for four dimensional spinors.

The possible dichotomy in the treatment of the helicity degrees of freedom has had an interesting 
effect in the derivation of sum rules based on the speculative assumption that one-loop 
quadratic divergencies cancel in the Standard Model (SM). As explained in Ref.~\cite{v8}, in DR 
the condition of cancellation of quadratic divergencies in one-loop tadpole diagrams 
is given by
\ini \label{e46b}
\mbox{Tr}\ \mathbbm{1}\ \sum_f m_f^2 = 3\ m_H^2 + (2\ m_W^2 + m_Z^2) (n-1)\, ,  
\fin
where the $f$ summation is over fermion masses and includes the color degree of freedom.
The factor $n-1$ reflects once more the helicity degrees of freedom of spin 1 bosons
in $n$ dimensions.
Eq.~\reff{e46b} leads also to the cancellation of all quadratic divergencies in the
one-loop contributions to the Higgs boson and fermion self-energies. Setting 
$n = \mbox{Tr}\ \mathbbm{1} = 4$, and neglecting the contributions of the lighter fermions
one obtains the Veltman-Nambu sum rule \cite{v6,v7}:
\ini \label{e47b}
m_H^2 = 4\ m_t^2 - 2\ m_W^2 - m_Z^2\, ,  
\fin

On the other hand, it was pointed out in Ref.~\cite{v8} that in DR Eq.~\reff{e46b}
with $n=4$ is not sufficient to cancel the remaining quadratic divergencies in the $W$
and $Z$ self-energies. Associating once more the one-loop quadratic divergencies with 
the $n=2$ poles, the cancellation of the residues in all cases ($f$, $H$, $W$, $Z$) takes
place when  $n=2$ is chosen. With $n = 2$ and $\mbox{Tr}\ \mathbbm{1} = 2$~\cite{v10}, 
this leads to the alternative sum rule \cite{v8}
\ini \label{e48b}
m_H^2 = 2\ m_t^2 - \frac{(2\ m_W^2 + m_Z^2)}{3}\, .  
\fin

Inserting the current values, $m_t = 174.3\ \mbox{GeV}$, $m_Z = 91.1875\ \mbox{GeV}$,
and  $m_W = 80.426\ \mbox{GeV}$ \cite{v11},  Eq.~\reff{e47b} and Eq.~\reff{e48b} lead
to the predictions $m_H = 317\ \mbox{GeV}$, and  $m_H = 232\ \mbox{GeV}$, 
respectively.
The current 95\% CL upper bound from the global fit to the SM is  
$m_H^{95} = 211\ \mbox{GeV}$ \cite{v11}, so that the above values are somewhat larger than the
range favored by the electroweak analysis. It will be interesting to see whether
these predictions ultimately bear any relation to reality!

\section{Effect of Interactions}

An important issue is what happens when interactions are taken into account.
The investigation of their effect on $\tll$ is an open and difficult one,
since vacuum matrix elements are factored out and then cancelled in the usual
treatment of Quantum Field Theory. As it is well-known, in the conventional framework,
interactions break the scale invariance of free-field theories
in the massless limit, a phenomenon referred to as the trace anomaly \cite{v12,v13,v14}.
One naturally expects that a similar phenomenon takes place in vacuum amplitudes,
an occurrence that would lead to the emergence of quartic divergencies. In this Section
we limit our analysis to two instructive examples.

We first discuss the question in the scalar theory with ${\cal L}_{int}=-(\lambda/4!)\phi^4$.
One readily finds that in ${\cal O}(\lambda)$ this interaction contributes $-(\lambda/8)\df^2(0)$
to $\rho$, which is quartically divergent. This result is obtained by either using
the familiar plane wave expansion involving annihilation and creation operators
and their commutation relations, or by the calculation of the relevant Feynman diagram,
which is a ``figure 8'' with the interaction at the intersection (see Fig.~2).
We note that $1/8=3/4!$ is the symmetry number for this diagram. However,
the mass in  Eq.~\reff{e12} is the unrenormalized mass $m_0$, since this is the
parameter that appears in the Lagrangian.
Writing $m^2_0 = m^2 - \Pi(m^2)$, where $\Pi$ is the self-energy,
Eq.~\reff{e12} generates a counterterm $-\Pi(m^2) i \df(0)/4$.
In ${\cal O}(\lambda)$ the only contribution to $\Pi(m^2)$ is the seagull diagram and equals
 $(\lambda/2)i \df(0)$, where $1/2$ is the symmetry number. Thus, the counterterm
$(\lambda/8)\df^2(0)$ exactly cancels the quartic divergence from the ``figure 8'' diagram!
The same cancellation occurs when regulator fields are present, since  Eq.~\reff{e17} 
involves just a linear combination of terms analogous to  Eq.~\reff{e12}!

Next we consider the example of QED in  ${\cal O}(e^2)$. Choosing the symmetric 
and explicitly gauge invariant version
of $T_{\mu \nu}$, one finds in n-dimensions:
\ini \label{e47}
{T^\lambda}_\lambda=\frac{n-4}{4}\ F_\mn\ F^\mn 
- (n-1)\bar{\psi} \left[ i \frac{\uplr{\slashb{D}}}{2}-m_0 \right]\psi 
+ m_0\ \bar{\psi}\ \psi  \, ,
\fin
where $F_{\mu \nu} = \partial_\mu A_\nu - \partial_\nu A_\mu $, 
$D_\mu =\partial_\mu +i e A_\mu$ is the covariant derivative
and $\psi$ and $A_\mu$ are the electron and photon fields.
It is easy to see that the insertion of $-(n-1) \bar{\psi} (i \uplr{\slashb{\partial}}/2 - m_0 ) \psi$ 
in the electron-loop,
corrected by the interaction in order $e^2$ (Fig.~3), cancels against the contribution
of $(n-1) e \bar{\psi} \slash{A} \psi$ corrected in  ${\cal O}(e)$ (see Fig.~4). This reflects
the validity of the equation of motion $(i \slashb{D} - m_0)\ \psi = 0$.

Next, we focus on the insertion of $[(n-4)/4] F_\mn F^\mn$ in the
photon line (Fig.~5). The fermion-loop integral with two external off-shell
photons of momentum $q$ is given by {\small
\ini \label{e48}
\frac{-i 8 e^2}{(2 \sqrt{\pi})^n}\ \mu^{(4-n)}\ \Gamma(2-n/2)\ 
(g_\mn q^2- q_\mu q_\nu)\ \int_0^1 dx\ x\ (1-x)\ [m^2-q^2 x (1-x)]^{\frac{n}{2}-2}\, ,
\fin}
where $m$ is the fermion mass. Closing the photon line and inserting the vertex
$[(n-4)/4] F_\mn F^\mn$ multiplies Eq.~\reff{e48} by
\bmath
\mu^{(4-n)}\ \int \frac{d^n q}{(2 \pi)^n}\frac{n-4}{2}\ q^2 g^\mn 
\left(\frac{-i}{q^2}\right)^2
\emath
and we obtain for the diagram of Fig.~5:
\bmath
{\cal A}_{DR} = \frac{4 i e^2}{(2 \sqrt{\pi})^n}\ (\mu^{4-n})^2\ (n-4)(n-1) \Gamma(2-n/2)\ 
\emath
\ini \label{e49}
\times \int_0^1 dx\ x\ (1-x) \int \frac{d^n q}{(2 \pi)^n} [m^2-q^2 x (1-x)]^{\frac{n}{2}-2}\, .
\fin
We note that $(1/q^2)^2$, arising from the two photon propagators in Fig.~5,
has been cancelled by two $q^2$ factors, one from  Eq.~\reff{e48}, the other 
from the $[(n-4)/4] F_\mn F^\mn$ vertex.

Performing a Wick rotation of the $q_0$ axis, and introducing 
\bmath
q_0 = i Q_0\, ,\quad \vec q = \vec Q\, , \quad 
d^n Q = \frac{\pi^{n/2}}{\Gamma(n/2)}\ Q^{2\left(\frac{n-2}{2}\right)} dQ^2\, ,\quad u = Q^2\, , 
\emath
we have {\small
\ini \label{e50}
{\cal A}_{DR} = - e^2 C(n) \int_0^1 dx \left[x (1-x)\right]^{\frac{n}{2}-1} 
\int_0^\infty du\ u^{\frac{n}{2}-1} \left[u+\frac{m^2}{x(1-x)}\right]^{\frac{n}{2}-2}\, ,
\fin }
where
\bmath
C(n) = \frac{4}{(4 \pi)^n}\ \frac{\Gamma(2-n/2)}{\Gamma(n/2)}(n-4)(n-1) (\mu^{4-n})^2\, . 
\emath
The $u$-integral in Eq.~\reff{e50} equals
\bmath
\left[ \Gamma(n/2) \Gamma(2-n)/\Gamma(2-n/2)\right]\ m^{2n-4} \left[x(1-x)\right]^{2-n}\, ,
\emath
and Eq.~\reff{e50} becomes {\small
\ini \label{e51}
{\cal A}_{DR} = - \frac{4\ e^2}{(4 \pi)^n} (n-4)(n-1) (\mu^{4-n})^2\ \Gamma(2-n)
 m^{2n-4} \int_0^1 dx \left[x(1-x)\right]^{1-n/2}\, .
\fin}
The remaining integral equals $B(2-n/2,2-n/2)=\Gamma^2(2-n/2)/\Gamma(4-n)$ and we find
\ini \label{e52}
{\cal A}_{DR} = \frac{16\ e^2\ m^4\ (\mu/m)^{2(4-n)} (n-1)\ \Gamma^2(3-n/2)}
{(4 \pi)^n (2-n)(3-n)(4-n)}  \, .
\fin
Eq.~\reff{e52} exhibits simple poles at $n=2, 3, 4$. According to the
rules of correspondence for two-loop diagrams, this indicates the presence
of quartic, quadratic, and logarithmic divergencies.

For clarity, we point out that potentially there is a counterterm diagram associated
with Fig.~5, in which the fermion loop is replaced by the insertion of the field 
renormalization vertex $- i \delta Z  F_\mn F^\mn /4$ in the closed photon loop.
However, such diagram, involving two vertices proportional to $F_\mn F^\mn$ and two 
photon propagators, leads to a result proportional to $i\int d^n q$,
which is imaginary in Minkowski space and
furthermore vanishes in DR. In particular, this also means that if the result for the
$k$-subintegration given in Eq.~\reff{e48} were expanded about $n=4$, the pole contribution
would cancel when the $q$ integration is performed. Since we need the full dependence on $n$
to determine the possible pole positions and the pole contribution from the
$k$-subintegration vanishes, in Eq.~\reff{e52} we have evaluated the full two-loop integral,
without expanding the $k$-subintegration  about $n=4$.

There remains the contribution of $m_0 \bar{\psi} \psi$, the last term in  Eq.~\reff{e47},
corrected by the interaction in order $e^2$. The superficial degree of divergence
in four dimensions of the corresponding two-loop diagram is trilinear, so that
one expects a quadratic divergence. Indeed, the DR calculation of the diagram shows a
pole at $n=3$ and a double pole at $n=4$ which, according to the rules of correspondence
for two-loop diagrams, indicate a quadratic divergence and logarithmic 
singularities proportional to ${\cal O} (m^4  \ln{(\Lambda^2/ m^2)})$ and 
${\cal O} (m^4  \ln^2{(\Lambda^2/ m^2)})$.

\section{Conclusions}

In this paper we discuss a number of issues related to the nature of the contributions of 
fundamental particles to the vacuum energy density. This problem is
of considerable conceptual interest since what may be called the physics 
of the vacuum is not addressed in the usual treatment of Quantum Field Theory.
On the other hand, it also represents a major unsolved problem since estimates of
these contributions show an enormous mismatch with the observed cosmological constant.

As a preamble to our analysis, in Section 2 we use an elementary argument
to derive rules of correspondence between the poles' positions in DR and ultraviolet cutoffs
in four dimensional calculations. In the case of quadratic divergencies, they coincide with
Veltman's dictum \cite{v6}, while they are extended here to quartic singularities.
A specific example of this correspondence is given at the one-loop level in Section 3. 

In Section 3 we address the Lorentz-covariant regularization of $\tmn$ in free-field theories,
starting from the elementary expressions for $\rho$ and $p$. Making use of
a mathematical identity, we are led to a covariant expression, which is immediately
confirmed by the direct evaluation of $\tll$. The regularization of this result is then
implemented in the PV, DR, and FR frameworks. In Section 4, we re-examine $\tmn$ on the basis of
the well-known expression for the vacuum expectation value of products of free fields, 
as well as one-loop Feynman vacuum diagrams,
with results that are consistent with those of Section 3. In Section 5, we consider two cases
involving interactions: $\lambda \phi^4$ theory in ${\cal O}(\lambda)$ and QED in ${\cal O}(e^2)$,
which require the examination of two-loop vacuum diagrams.

Our general conclusion, based on Lorentz covariance and the scale invariance 
of free-field theories in the massless limit, as well as consistency with the rules of
correspondence applied to $\tll$, is that quartically divergent contributions to $\rho$
are absent in the case of free fields.

At first hand, the notion that free photons do not contribute to $\rho$ may
seem strange. However, we point out that this immediately follows from 
Eq.~\reff{e47}, which tells us
that for free photons $({T^\lambda}_\lambda)_\gamma = 0$ in four dimensions. This
implies $({t^\lambda}_\lambda)_\gamma = 0$ and, using Eqs.~(\ref{e2}, \ref{e3}), 
$\rho_\gamma = 0 $! In more pictorial language: $({T^\lambda}_\lambda)_\gamma = 0$
implies $\rho_\gamma = 3 p_\gamma$, the equation of state of a photon gas but,
in the vacuum case, Eq.~\reff{e2} tells us that $\rho^{vac} = - p^{vac}$
for any field. The only way of satisfying the two constraints in the vacuum case
is  $\rho^{vac}_\gamma = p^{vac}_\gamma = 0$.

As pointed out in Section 3, in the case of free fields the same conclusion was 
recently advocated in the interesting work of E.~Kh.~Akhmedov \cite{v5} on the basis
of a less complete line of argumentation, and without examining the effect of 
interactions.

When interactions are turned on, as illustrated in the QED case in Section 5, our 
conclusion is that quartic divergencies generally emerge.
Thus, in some sense there is a parallelism between the analysis of vacuum amplitudes 
and conventional Quantum Field Theory. 
Free-field theories are scale invariant in the massless limit and, 
according to our interpretation, this partial symmetry protects the theory
from the emergence of quartic
divergencies. However, in the presence of interactions, the symmetry is broken even
in the massless limit and consequently such singularities generally arise. On the 
other hand, it is worthwhile to recall that ${T^\lambda}_\lambda$ becomes 
a soft operator even in the presence of interactions under the speculative assumption
that the coupling constants are zeros of the relevant $\beta$-functions \cite{v13,v14}.

From the point of view of formal renormalization theory, the presence of the highly
divergent expressions encountered in the study of vacuum amplitudes does not present
an insurmountable difficulty. For instance, in the PV approach discussed in Section 3
with a sufficiently large number of regulator fields, the coefficients of the
free-field divergencies are undefined and in principle can be chosen to cancel the
corresponding singularities emerging from interactions. More generally,
the $\lambda$ constant that appears in Einstein's equation can be adjusted to
cancel such singularities. As emphasized by several authors \cite{v1} the crisis resides
in the extraordinarily unnatural fine-tuning that these cancellations entail.
  
From a practical point of view, the conclusions in the present paper hardly affect
the cosmological constant problem: clearly, it makes very little difference phenomenologically
whether the mismatch is 123 or 120 orders of magnitude! On the other hand, they place the origin
of the problem on a different conceptual basis.

The simplest framework in which quartic divergencies cancel remains supersymmetry
since it implies an equal number of fermionic and bosonic degrees of freedom.
In some effective supergravity theories derived from four dimensional superstrings,
with broken supersymmetry, it is possible to ensure also the cancellation of the 
${\cal O}(m^2 \Lambda^2)$ terms in the one-loop effective potential \cite{v15}.
In such scenarios, the ${\cal O}(m^4)$  terms become ${\cal O}(m_{3/2}^4)$,
where $m_{3/2}$, the gravitino mass, is associated with the scale of supersymmetry 
breaking. Assuming $m_{3/2} ={\cal O}(1\ \mbox{TeV})$ this leads to the rough
estimate $\rho = {\cal O}(\mbox{TeV}^4)$ mentioned in the Introduction.

\section*{Acknowledgments}

The authors are greatly indebted to Steve Adler, Martin Schaden, Georgi Dvali, 
and Massimo Porrati for valuable discussions and observations. 
One of us (A.S.) would also like to thank Georgi Dvali for 
calling Akhmedov's paper \cite{v5} to his attention, after completion of an earlier and
less detailed version of this paper, and 
Michael Peskin, Glennys Farrar, Antonio Grassi, Peter van Nieuwenhuizen, Robert Shrock,
and William Weisberger for useful discussions.
This work was supported in part by NSF Grant No.~PHY-0070787.

\appendix
\section*{Appendix A}
\begin{appendletterA}

In this Appendix we analyze Eq.~\reff{e20}, which is the PV-regularized version of 
Eq.~\reff{e12}. Choosing the minimum number $N=3$ of regulator fields,
the constants $C_j$ ($j=1,2,3$) can be expressed in terms of the $m^2$ and the $M_j^2$
($j=1,2,3$) by solving  Eq.~\reff{e16} for $p=0,1,2$. This leads to
\ini
C_1=\frac{- M_2^2 M_3^2 + m^2 (M_2^2 + M_3^2) - m^4} {(M_2^2 - M_1^2) (M_3^2 - M_1^2)}\, . 
\fin
$C_2$ is obtained from $C_1$ by applying the cyclic permutation (1 2 3), while
$C_3$ is obtained from $C_2$ by means of the same permutation.
Focusing on the quartic divergencies, we neglect the terms proportional
to $m^2$ and $m^4$ and,since $\nu^2$ in Eq.~\reff{e20} is arbitrary, we
choose $\nu^2 = M_3^2$. Then the quartically divergent terms in Eq.~\reff{e20}
can be cast in the form
\ini \label{a2}
(\tmn)_{PV}(m^2=0)=\frac{g_\mn}{64 \pi^2} \frac{M_1^2 M_2^2 M_3^2}{M_2^2 - M_1^2}
\left[ f(\frac{M_3^2}{M_1^2}) -  f(\frac{M_3^2}{M_2^2})\right]\, ,
\fin
where
\ini
f(x) = \frac{\ln{x}}{x-1}\, .
\fin
We note that $f(x)$ is positive definite for all $x \ge 0$ while its derivative
\ini
f'(x) =  \frac{1}{x-1} \left[\frac{1}{x}-\frac{\ln{x}}{x-1}\right]
\fin
is negative definite. Thus, $f(x)$ is a positive definite and decreasing 
function of its argument. Consider now the case $M_2^2 > M_1^2$. Then
$M_3^2/M_1^2 > M_3^2/M_2^2$ and the expression between square brackets
is negative. Since $(M_2^2 - M_1^2) > 0$, we
conclude that the coefficient of $g_\mn$ in Eq.~\reff{a2} is negative for any
value of $M_3^2$. If  $M_1^2 > M_2^2$, $M_3^2/M_1^2 < M_3^2/M_2^2$,
the square bracket is positive but $(M_2^2 - M_1^2) < 0$,
so that we reach the same conclusion.
Thus, for all possible values of the regulator masses $M_j^2$,
the coefficient of $g_\mn$ in Eq.~\reff{e20} is negative definite which, as explained 
in Section 3, is physically unacceptable.
Analogous arguments show that for all $M_j^2$, the contributions 
of ${\cal O} (m^2 M_j^2)$ and ${\cal O} (m^4)$
in the cofactor of $g_\mn$ are positive and negative definite,
respectively. In the limit  $M_3 \to M_2 \to M_1 = \Lambda$,  Eq.~\reff{a2}
greatly simplifies and reduces to the first term in Eq.~\reff{e21}.

As mentioned in Section 3, one possible solution of the sign
problem is to subtract the offending  ${\cal O} (M^4)$ contributions.
Another possibility is to consider $N \ge 4$ regulator fields.
In that case Eq.~\reff{e16} for $p=0,1,2$ are not sufficient to determine 
the  $C_j$ ($j=1,\ldots,N$) in terms of the masses. As expected, we have checked 
that the coefficient of the  ${\cal O} (M^4)$ term in  Eq.~\reff{e20} becomes undetermined
while still satisfying the three relations of Eq.~\reff{e16}, so that it may be chosen 
to be positive or, for that matter, zero.
The last possibility would naturally follow by invoking the scale invariance of free
field theories in the massless limit and would conform with the analysis
based on DR. As mentioned in Section 3, in the $N \ge 4$ solution of the problem,
the coefficient of the leading  ${\cal O} (m^2 \Lambda^2)$ is also undefined,
which is not a satisfactory state of affairs!
\end{appendletterA}

\section*{Appendix B}
\begin{appendletterB}

In this Appendix we apply DR and the rules of correspondence to integrals
that occur in the evaluation of the one-loop effective potential
in the $\lambda \phi^4$ theory \cite{v18}:
\ini \label{b1}
V(\phi_c) = -\frac{i\ \mu^{4-n}}{2} \int \frac{d^n k}{(2 \pi)^n} 
\ln{ \left( \frac{k^2-m^2-\frac{1}{2}\lambda \phi_c^2 + i \epsilon}{k^2 - m^2+ i \epsilon} 
\right) }\, .
\fin
We recall that in the path integral formalism, the denominator in the argument
of the logarithm arises from the normalization of the generating functional $W[J]$,
namely $W[0]=1$.

We consider the integral:
\ini \label{b2}
K = -\frac{i\ \mu^{4-n}}{2} \int \frac{d^n k}{(2 \pi)^n}  \ln{(k^2 - c+ i \epsilon)} \, , 
\fin
which can be obtained from
\ini
L = -\frac{i\ \mu^{4-n}}{2} \int \frac{d^n k}{(2 \pi)^n} (k^2 - c+ i \epsilon)^{\alpha} 
\fin
by differentiating with respect to $\alpha$ and setting $\alpha = 0$. The
last integral is given by
\ini
L = \frac{\mu^{4-n}}{2 (2 \sqrt{\pi})^n}\ (-1)^{\alpha}\ c^{\frac{n}{2}+\alpha}\  
\frac{\Gamma(-\alpha-\frac{n}{2})}{\Gamma(-\alpha)}\, . 
\fin
Since $1/\Gamma(0) = 0$, the only non-vanishing contribution to $K$
involves the differentiation of $\Gamma(-\alpha)$. Thus
\ini \label{b5}
K = - \frac{\mu^{4-n}}{2 (2 \sqrt{\pi})^n}\  c^{\frac{n}{2}}\ \Gamma(-\frac{n}{2}) 
\fin
where we have employed $\lim_{\alpha \to 0}{\psi(-\alpha)/\Gamma(-\alpha)} = -1$.
We see that $K$ contains poles at $n=0,2,4$ which, according to the rules of correspondence,
indicate quartic, quadratic and logarithmic ultraviolet singularities. However, Eq.~\reff{b1}
involves the difference of two integrals of the $K$ type and we find
\ini \label{b6}
V(\phi_c) = -\frac{\mu^{4-n}}{2 (2 \sqrt{\pi})^n} \left[ \left(m^2 + \frac{\lambda \phi_c^2}{2} \right)^{n/2}
-  (m^2)^{n/2} \right]\ \Gamma(-\frac{n}{2})\, . 
\fin
Clearly, the residue of the $n=0$ pole cancels in Eq.~\reff{b6} and the leading singularity
is given by the $n=2$ pole:
\ini \label{b7}
V(\phi_c) = \frac{\mu^2}{8 \pi} \frac{\lambda\ \phi_c^2}{(2-n)} + \cdots \, ,
\fin
which corresponds to a quadratic divergence. 
This conforms with the result for the leading singularity obtained by
expanding  Eq.~\reff{b1} in powers of $\lambda \phi_c^2$.
Moreover, if one approaches the pole from below, as discussed after Eq.~\reff{e34}, the
signs also coincide! As it is well-known, the quadratic and logarithmic singularities in
Eq.~\reff{b6} are cancelled by the $\delta m^2$ and $\delta \lambda$ counterterms \cite{v18}. 

The $K$ integral with $c=m^2$ is also interesting because in some formulations it is directly
linked to the vacuum energy density contribution from free scalar fields \cite{v19}.
In order to obtain a four-dimensional representation of Eq.~\reff{b2}, we introduce a Feynman
regulator $[\Lambda^2/(\Lambda^2-k^2 - i \epsilon)]^3$ and perform a Wick rotation, which leads
to the Euclidean-space expression 
\ini \label{b8}
K = \frac{1}{32\pi^2} \int_0^\infty du\ u \ln{\left(\frac{u + m^2}{\sigma^2}\right)}
\left(\frac{\Lambda^2}{\Lambda^2+u}\right)^3  \, . 
\fin
In order to give mathematical meaning to the logarithm, in Eq.~\reff{b8} we have
introduced a squared-mass scale $\sigma^2$ which, for the moment, is unspecified.
Evaluating Eq.~\reff{b8}, we have
\ini \label{b9}
K = \frac{1}{64\pi^2}\left\{ \Lambda^4 \left[ \ln{\left(\frac{\Lambda^2}{\sigma^2}\right)} 
+ 1 \right] + m^2\Lambda^2-m^4 \left[ \ln{\left(\frac{\Lambda^2}{m^2}\right)} 
- 1 \right]
\right\} \, , 
\fin
where we have neglected terms of  ${\cal O} (m^2 / \Lambda^2)$.
In a free-field theory calculation, one expects the answer to depend on $m^2$ and $\Lambda^2$,
as we found, for instance, in Eq.~\reff{e21} and Eq.~\reff{e27}. By arguments
analogous to those explained at the end of Section 3,
one finds that $\sigma^2$ cannot be identified with $m^2$. In fact,
with $c = m^2$,  Eq.~\reff{b5} is proportional to $(m^2)^{n/2}$.
Differentiating with respect to $m^2$, the $n=0$ pole in  Eq.~\reff{b5} cancels.
This implies that terms of  ${\cal O} (\Lambda^4 \ln{(\Lambda^2 / m^2)})$ cannot be present,
since otherwise contributions of ${\cal O} (\Lambda^4 )$ would survive the differentiation 
with respect to $m^2$.
An attractive idea to fix  $\sigma^2$ is to invoke symmetry considerations. In particular,
since according to the arguments of this paper the terms of ${\cal O} (\Lambda^4 )$ 
violate the scale invariance of free field theories in the massless limit,
we may choose $\sigma=\sqrt{e}\ \Lambda$ to eliminate such contributions.
In that case, Eq.~\reff{b9} reduces to our previous result in  Eq.~\reff{e27}, obtained
by more elementary and transparent means!
Correspondingly, in the DR version of $K$, the $n=0$ pole $\mu^4 (4-n)/4n$
may be removed in order to conform with the scale invariance of free field 
theories in the massless limit. This can be achieved by appending an $n/4$ normalization factor
to the r.h.s. of  Eq.~\reff{b5}, in which case the rules of correspondence
between the DR and four-dimensional calculations reduce precisely to  Eqs.~(\ref{e33}, \ref{e34}).

In summary, aside from the fact that the derivation of Eq.~\reff{e12} and the calculation
of  Eq.~\reff{e27} are particularly simple, they offer the additional advantage that they 
explicitly exhibit the partial scale invariance of free-field theories.

\end{appendletterB}

\newpage

\begin{figure}[ht]
\begin{center}
\begin{picture}(300,100)(0,0)
\DashCArc(150,50)(40,180,540){3}
\Text(110,50)[]{\inse}
\end{picture}
\end{center}
\caption{One-loop vacuum amplitudes. The cross indicates the insertion of
the trace ${T^\lambda}_\lambda$ of the energy-momentum tensor. The dashed
line represents a scalar, spinor or massive vector particle (see Section~4).}
\end{figure}
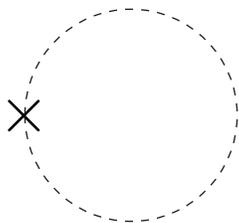

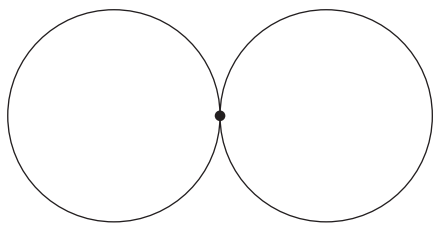
\begin{figure}[ht]
\begin{center}
\begin{picture}(300,100)(0,0)
\CArc(110,50)(40,0,360)
\CArc(190,50)(40,0,360)
\Vertex(150,50){2}
\end{picture}
\end{center}
\caption{Two-loop vacuum amplitude in $\lambda \phi^4$ theory. (See Section~5).}
\end{figure} 

\begin{figure}[ht]
\begin{center}
\begin{picture}(300,100)(0,0)
\ArrowArc(150,50)(40,90,180)
\ArrowArc(150,50)(40,180,270)
\ArrowArc(150,50)(40,270,90)
\Photon(150,10)(150,90){4}{10}
\Text(110,50)[]{{\inse}}
\Vertex(150,10){2}
\Vertex(150,90){2}
\end{picture}
\end{center}
\caption{Two-loop vacuum amplitude in QED. The cross represents the insertion of 
$-(n-1) \bar{\psi} (i \uplr{\slashb{\partial}}/2 - m) \psi$ (Section~5).}
\end{figure}

\begin{figure}[ht]
\begin{center}
\begin{picture}(300,100)(0,0)
\ArrowArc(150,50)(40,0,180)
\ArrowArc(150,50)(40,180,360)
\Photon(110,50)(190,50){4}{10}
\Text(110,50)[]{\inse}
\Vertex(190,50){2}
\end{picture}\caption{Two-loop vacuum amplitude in QED. The cross represents the 
insertion of $(n-1) e \bar{\psi} \slash{A} \psi$ (Section~5).}
\end{center}\end{figure}

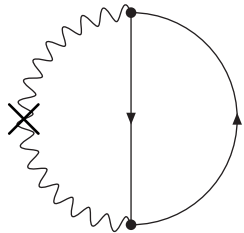
\begin{figure}[ht]
\begin{center}
\begin{picture}(300,100)(0,0)
\PhotonArc(150,50)(40,90,270){3}{16}
\ArrowArc(150,50)(40,270,90)
\ArrowLine(150,90)(150,10)
\Text(110,50)[]{\inse}
\Vertex(150,10){2}
\Vertex(150,90){2}
\end{picture}\caption{Two-loop vacuum amplitude in QED. The cross represents the 
insertion of  $[(n-4)/4] F_\mn F^\mn$ (Section~5).}
\end{center}
\end{figure}

\end{document}